\input amssym.def
\input amssym
\documentstyle[12pt]{article} 
\newcommand{\doublespace}{
   \renewcommand{\baselinestretch}{1.5}
   \large\normalsize}

\def \Z{\Bbb Z}
\def \C{\Bbb C}
\def \R{\Bbb R}
\def \Q{\Bbb Q}

\def \wt{{\rm wt}}

\def \Res{{\rm Res}}
\def \QRes{{\rm QRes}}
\def \End{{\rm End}}
\def \E{{\rm End}}
\def \Ind {{\rm Ind}}
\def \Irr {{\rm Irr}}
\def \Aut{{\rm Aut}}

\def \<{\langle} 
\def \>{\rangle} 

\def \a{\alpha }

\def \l{\lambda }
\def \L{\Lambda }
\def \g{\gamma}
\def \b{\beta }

\def \c{\chi}
\def \ch{\chi}

\doublespace
\begin{document}
\newtheorem{th1}{Theorem}
\newtheorem{th}{Theorem}[section]
\newtheorem{prop}[th]{Proposition}
\newtheorem{coro}[th]{Corollary}
\newtheorem{lem}[th]{Lemma}
\newtheorem{rem}[th]{Remark}
\newtheorem{de}[th]{Definition}

\begin{center}
{\Large {\bf On Quantum Galois Theory}} \\
\vspace{0.5cm}
Chongying Dong\footnote{Supported by NSA grant MDA904-92-H-3099, NSF grant 
DMS-9303374, a Regent's Junior Faculty Fellowship of the University of California and a research grant from the Committee on Research, UC Santa Cruz.} \ \ \  and \ \ \ Geoffrey Mason\footnote{Supported by NSF grant DMS-9122030 and a research grant from the Committee on Research, UC Santa Cruz.}
\\
Department of Mathematics, University of
California, Santa Cruz, CA 95064
\end{center}

\hspace{1.5 cm}

\section{Introduction}
The goals of the present paper are to initiate a program to systematically 
study and rigorously establish what a physicist might refer to as the 
``operator content of orbifold models.'' To explain what this might mean,
and to clarify the title of the paper, we will assume that the reader
is familiar with the algebraic formulation of 2-dimensional CFT
in the guise of vertex operator algebras (VOA), see [B], [FLM]
and [DM] for more information on this point.

In the paper [DVVV], several ideas are proposed concerning the structure
of a holomorphic orbifold. In other words, if $V$ is a holomorphic 
VOA and if $G$ is a finite group of automorphisms of $V,$ then the sub VOA
$V^G$ of $G$-invariants is itself a VOA and the subject of [DVVV]
is very much concerned with speculation on the nature of the $V^G$-modules.

It turns out to be more useful $-$ at least for purpose of inductive proofs
$-$ to take $V$ to be a {\it simple} VOA. We will then see 
that $V^G$ is also simple whenever $G$ is a finite
group of automorphisms of $V.$ One consequence of our main results is the 
following:
\begin{th1}\label{tt1} Let $V$ be a simple VOA and $G$ a finite and 
faithful group of
automorphisms of $V.$ Assume that $G$ is either abelian or dihedral.
Then there is a bijection between the subgroups of $G$ and the 
sub VOAs of $V$ which contain $V^G$ defined by the map $H\mapsto V^H.$
\end{th1}

This partially explains the title of the paper. We have no reason to believe
that Theorem \ref{tt1} is not true for any finite group $G,$ and indeed we
can prove the analogous result for various  classes of groups other than 
just those listed in Theorem \ref{tt1}.

There are other Galois correspondences which may well be relevant to the
present context. Three that come to mind are the theory of $G$-covering
spaces, and Galois theory in the context of von Neumann algebras [J]
and Hopf algebras [Mo]. Indeed Iiyori and Yamada have taken in [IY]
an approach to results related to Theorem \ref{tt1} with von
 Neumann algebras very much in mind.

Our own approach yields Theorem \ref{tt1} as essentially a consequence of 
results concerning the structure of $V$ as a $V^G$-module. To explain these,
let $V$ be a simple VOA and $G$ a finite and faithful 
group of automorphisms of $V,$ and
let $\Irr(G)$ denote the set of simple characters $\chi$ of $G.$ Now
as ${\C}G$-module, each homogeneous space $V_n$ of $V$ is of finite dimension
and so there is a direct sum decomposition of $V$ into graded subspaces
\begin{equation}\label{1.1}
\displaystyle{V=\oplus_{\chi\in \Irr(G)}V^{\chi}}
\end{equation}
where $V^{\chi}$ is the subspace of $V$ on which $G$ acts according to the 
character $\chi.$ In other words, if $M_{\chi}$ is the simple ${\C}G$-module
affording $\chi$ then $V^{\chi}$ is the $M_{\chi}$-homogeneous subspace
of $V$ in the sense of group representation theory. 

\begin{th1}\label{t2} Each $V^{\chi}$ is non-zero.
\end{th1}

This innocuous-looking assertion is, in a sense, the first assumption of [DVVV]. 
Even in the case that $G$ is abelian it can be used to substantially 
strengthen results in [IY]. Our proof uses a duality argument which 
generalizes a result in [DL].

In order to understand $V^{\chi}$ it is useful to have the notion
of a $G$-{\em graded vertex operator algebra}. We will introduce 
the formal definitions in Section 2, but essentially it is a generalization of a VOA in which the underlying vector space carries a $G$-grading
and the second term of the Jacobi identity includes a certain commutator. Such objects arise naturally in the representation theory of orbifold models
in much the same way that $G$-graded Clifford systems arise in group 
representation theory (cf. [CR]). A typical example of a $G$-graded VOA
is the tensor product $\C G\otimes V$ of the group algebra $\C G$ and 
a VOA $V.$ $G$-graded VOAs have an accompanying notion of representation.

We can now state the main results of the present paper:

\begin{th1}\label{t3} Suppose that $V$ is a simple VOA and that $G$ is a
finite and faithful solvable group of automorphisms of $V.$ Then the following
hold:

(i) For $\chi\in \Irr(G),$ each $V^{\chi}$ is a simple module for
the $G$-graded VOA $\C G\otimes V^G$ of the form
\begin{equation}\label{1.2}
V^{\chi}=M_{\chi}\otimes V_{\chi}
\end{equation}
where $M_{\chi}$ is the simple $\C G$-module affording $\chi$ and where
$V_{\chi}$ is a simple $V^G$-module.

(ii) The map $M_{\chi}\mapsto V_{\chi}$ is a bijection from the set
of simple $\C G$-modules to the set of (inequivalent) simple $V^G$-modules
which are contained in $V.$
\end{th1}

Combining Theorems \ref{t2} and \ref{t3}, we find that, at least for 
solvable groups, there is a decomposition of $V$ into simple $V^G$-modules
of the form
$$V=\oplus_{\chi\in \Irr(G)}(\dim \chi)V_{\chi}.$$
This is exactly the result predicted in [DVVV]. 

Introduce the category 
$G$-Mod of (left) $\C G$-modules and the category ${\cal V}(G)$ of 
$V^G$-modules  whose objects are direct sums of simple $V^G$-modules which are
contained in $V.$ Since $V$ is completely reducible as $V^G$-module by
Theorem \ref{t3} (at least
if $G$ is solvable), then ${\cal V}(G)$ may also be described as the category
whose objects are $V^G$-submodules of direct sums of copies of $V.$ Then
the map in part (ii) above induces an equivalence of categories 
$$G\!-\!{\rm Mod}\stackrel{\phi_G}{\longrightarrow}{\cal V}(G).$$

\begin{th1}\label{t4} Let $V$ be a simple VOA, and let $G$ be a finite
and faithful solvable group
of automorphisms of $V$ with $H\leq G.$ Then there is
a commuting diagram
$$\begin{array}{ccc}
\hspace{1 cm} G\!-\!{\em Mod} &\stackrel{\phi_G}{\longrightarrow} & {\cal V}(G)
\hspace{1 cm}  \\
\Ind_H^G\uparrow &  & \uparrow \QRes_H^G\\
\hspace{1 cm}H\!-\!{\em Mod} &\stackrel{\phi_H}{\longrightarrow} & {\cal V}(H)
\hspace{1 cm}
\end{array}$$
Here, $\Ind_H^G$ is the usual induction functor for group 
representations, while $\QRes_H^G$ is the functor which restricts $V^H$-modules
to $V^G$-modules via the containment $V^G\subset V^H.$
\end{th1}

Let us remark that it is well-known that in order to understand all of the
simple $V^G$-modules one must consider not only $V$ itself but also
the simple $g$-twisted $V$-modules (cf. [DM], [Ma]). Under appropriate 
assumptions 
there are theorems corresponding to the foregoing which hold in this 
more general setting. To state and prove these would involve a profusion of 
notation though the ideas involved are the same. 
We thus leave the formulation of such results to the interested 
reader.

The rest of the paper is arranged as follows: in Section 2 we review the
general ideas involved in VOA theory, including the above-mentioned
$G$-graded VOAs. In Section 3 we prove, among other things,
Theorem \ref{t2}, and Section 4 contains
a result on generation of $V$-modules. 
This is applied in Section 5 
to deal with Theorems \ref{tt1} and \ref{t3} for the case of $G$ abelian,
and in Section 6 to deal with Theorem \ref{t4} and the more general
version of Theorem \ref{t3}. Theorem 1 for dihedral group occupies Section 7.
\newpage

\section{Vertex operator algebras and their modules}
\setcounter{equation}{0}

We begin with the definition of $G$-graded VOA. So let $G$ be a finite group.
A $G$-graded VOA is a $\C$-vector space carring a grading by $G\times\Z:$
\begin{equation}\label{2.1}
V=\coprod_{x\in G\times\Z}V_x
\end{equation}
such that $\dim V_{x}$ is finite for all $x\in G\times\Z$ and
 $ V_{x}=0$ for $x=(g,n)$ and $n << 0.$ 
Furthermore certain axioms are required. To explain these, let
\begin{equation}\label{2.2'}
\begin{array}{c}
\displaystyle{
V_g=\coprod_{n\in\Z}V_{(g,n)},\ \ \ g\in G}\\
\displaystyle{
V_n=\sum_{g\in G}V_{(g,n)},\ \ \ n\in \Z.}
\end{array}
\end{equation}
Then we require the existence of distinguished vectors
$v_g\in V_{(g,0)},$ ${\bf 1}\in V_{(1,0)}$ (vacuum),
$\omega\in V_{(1,2)}$ (Virasoro element) and  
a linear injection
$$
\begin{array}{l}
V \to (\mbox{End}\,V)[[z,z^{-1}]]\\
v\mapsto Y(v,z)=\displaystyle{\sum_{n\in\Z}v_nz^{-n-1}}\ \ \ \  (v(n)\in
\mbox{End}\,V)
\end{array}$$
such that  
the following hold for $u\in V_g, v \in V_h$ and $m,n\in\Z:$
\begin{eqnarray}
& &u_mv\in V_{gh}  \ \ {\rm and}\ \ u_nv=0\ \ \mbox{for}\ \  n\ \ 
\mbox{sufficiently large};\\
& &Y({\bf 1},z)=1;\label{ident}\\
& &Y(v,z){\bf 1}\in V[[z]]\ \ \ \mbox{and}\ \ \ \lim_{z\to
0}Y(v,z){\bf 1}=v; \label{rident}
\end{eqnarray}
The components of the vertex operator $Y(\omega,z)=\sum_{n\in\Z}L_nz^{-n-2}$
generate a copy of the Virasoro algebra represented on $V$ with
central charge $c\in\Q;$
\begin{eqnarray} 
& &L(0)v=nv \ \ \ \mbox{for}\ \ \ v\in V_n; \label{3.40}\\
& &\frac{d}{dz}Y(v,z)=Y(L(-1)v,z);\label{3.41}
\end{eqnarray}
Each $v_g$ is an {\em invertible} element of $V$ i.e., $Y(v_g,z)=
(v_g)_{-1}$ is an invertible operator on $V$ (we often identify $v_g$ 
with the corresponding operator on $V$);
\begin{equation}\label{2.2}
\begin{array}{c}
\displaystyle{z^{-1}_0\delta\left(\frac{z_1-z_2}{z_0}\right)
Y(u,z_1)Y(v,z_2)-[v_g,v_h]z^{-1}_0\delta\left(\frac{z_2-z_1}{-z_0}\right)
Y(v,z_2)Y(u,z_1)}\\
\displaystyle{=z_2^{-1}\delta
\left(\frac{z_1-z_0}{z_2}\right)
Y(Y(u,z_0)v,z_2)}
\end{array}
\end{equation}
(Jacobi identity) where 
$\delta\left(\frac{z_1-z_2}{z_0}\right)$ etc. is interpreted in the usual
way (cf. [FLM]) and where $[v_g,v_h]=v_gv_hv_g^{-1}v_h^{-1}.$
We denote the $G$-graded VOA just defined by
$(V,Y,G,{\bf 1},\omega)$ (or briefly, by $V$).

If $V$ is a $G$-graded VOA, a $V$-{\em module} is a $\Q$-graded 
vector space $M=\coprod_{n\in\Q}M_{n}$ together with
a linear map $V\rightarrow (\End M)[[z,z^{-1}]],$ $v\mapsto Y_M(v,z)
=\sum_{n\in\Z}v_nz^{-n-1}$ such that ``all the axioms in the definition
of $G$-graded vertex operator algebra which make sense hold.''

\begin{rem}\label{r2.1} (i) If $G=1$ then a $G$-graded VOA is simply a VOA.

(ii) More generally, if $v_1$ is a scalar operator then $V_1$ is a VOA and
the spaces $V_g$ are (isomorphic) $V_1$-modules.

(iii) If $G$ is a finite group, $\alpha$ a 2-cocycle i.e., 
$\alpha\in C^2(G,\C^*),$ and $V$ is a VOA then we can turn 
$(\C G)_{\alpha}\otimes V$ into a $G$-graded VOA. Here,
$(\C G)_{\alpha}$ is the $\alpha$-{\em twisted} group algebra (cf. [CR],
Section 8B), and one sets
$$Y(g\otimes v,z)=g\otimes Y(v,z)$$
$$v_g=g\otimes {\bf 1}.$$

(iv) If $V$ is a $G$-graded VOA and $V_1$ is a {\em simple} VOA (i.e., 
the only subspaces of $V_1$ invariant under all $v_n$ for $v\in V_1,$
$n\in\Z$ are 0 and $V_1$) then we can show  that
$V$ is necessarily a tensor product of the type in (iii) for
some choice of $\alpha.$
\end{rem}

There are a number of immediate and useful consequences of these axioms,
for which we refer the reader to [FLM], [FHL] and elsewhere.

An {\em automorphism} of an ordinary vertex operator algebra $V$ 
is an invertible linear map $g: V\to V$ satisfying:
\begin{equation}\label{2.11}
\begin{array}{c}
\displaystyle{
gY(v,z)g^{-1}=Y(gv,z)}\\
\displaystyle{
g{\bf 1}={\bf 1},\ \ g\omega=\omega.}
\end{array}
\end{equation}
In other words, we have $(gv)_n=gv_ng^{-1}$ for all $v\in V$ and $n\in\Z.$
Two modules $M_1$ and $M_2$ for $V$ are isomorphic if there exists an 
invertible
linear map $f:M_1\to M_2$ such that $fY_{M_1}(v,z)=Y_{M_2}(v,z)f$ for $v\in V.$
We shall use the notation $M_1\simeq M_2$ to indicate that
$M_1$ and $M_2$ are isomorphic $V$-modules.
There is the obvious notion of simple $V$-module, and in particular
$V$ is called {\em simple}  if it is itself a simple $V$-module. 

The theory  of the Zhu algebra $A(V)$ [Z] will be important to us, and we next recall some of the relevant facts. See [Z] for more details.

Let $V$ be a VOA, with $u,v\in V$ and $u$ homogeneous. (This means that $u\in V_n$ for some $n\in\Z.$ We denote $n$ by $\wt u,$ the weight of $u.$)
Define
\begin{equation}\label{2.12}
u\circ v=
\Res_{z}Y(u,z)\frac{(1+z)^{\wt u}}{z^2}v
\end{equation}
\begin{equation}\label{a2}
u*v=\Res_{z}Y(u,z)\frac{(1+z)^{\wt u}}{z}v.
\end{equation}
Let $O(V)$ be the subspace of $V$ spanned
by the product $u\circ v$ extended linearly to $V,$ and define  
$A(V)=V/O(V).$ 

\begin{th}\label{t2.1} [Z] The following hold:

(i) The product $*,$ extended linearly to $V,$ 
induces the structure of associative algebra on $A(V).$

(ii)  Let $M=\coprod_{n\geq 0}M_{n+n_0}$ be a $V$-module
with $M_{n_0}\ne 0.$ Then for $v\in V,$ the
operator $o(v)=v_{\wt v-1}$ leaves $M_{n_0}$ invariant and the map
$v\to o(v)$ is a  
representation of $A(V)$ on $M_{n_0}.$

(iii) If $M$ is a simple $V$-module then $M_{n_0}$ is  
a simple $A(V)$-module, and the map $M\mapsto M_{n_0}$ is an injection
from simple $V$-modules to simple $A(V)$-modules.
\end{th}

Suppose next that $V$ is a VOA, $G$ a group of automorphisms, and ${\cal M}$
is the set of simple $V$-modules. There is a permutation action of $G$ on 
${\cal M}$ defined as follows: if $M\in{\cal M},$ $Y_M(v,z)$
the vertex operator of $v\in V$ operating on $M,$ and $g\in G,$ then $g\circ M$ is defined to be the space
$M$ and the corresponding vertex operator is
$${}_gY_M(v,z)=Y_M(gv,z).$$
It is readily verified that indeed $g\circ M\in{\cal M}.$
\newpage

\section{Proof of Theorem 2}

\setcounter{equation}{0}

We establish Theorem \ref{t2} in this section. We start
by recalling some facts about duality. Let $V$ be a VOA and 
$W=\coprod_{n\geq n_0}W_n$ a $V$-module.
Let $W_n^*$ be the dual space of $W_n$ and 
$W'=\coprod_{m\geq m_0}W_m^*$ the restricted dual of $W.$ We denote by
$\langle \cdot,\cdot\rangle: W'\times W\to \C$
the restricted paring such that
$\langle W_n^*,W_m\rangle=0$ unless $m=n.$

Let $R$ be the subring of $\C(z_1,z_2)$ 
obtained by localizing $\C[z_1,z_2]$ at the set
consisting of products of non-zero homogeneous polynomials of degree 1. 
For each of the two orderings $(i_1,i_2)$ of the set $\{1,2\}$ there
is an injective ring map
$$\iota_{i_1i_2}: R\to \C[[z_1,z_1^{-1},z_2,z_2^{-1}]]$$
by which an element $(az_1+bz_2)^{-1}\in R$ is
 expanded in nonnegative integral powers of $z_{i_2}.$ Then we have

\noindent (i) {\bf Rationality}: Let $u,v\in V, w
\in W, w'\in W'.$ Then there is $f\in R$ of the form 
$f(z_1,z_2)=h(z_1,z_2)/z_1^rz_2^{s}(z_1-z_2)^t$ with
$h(z_1,z_2)\in \C[z_1,z_2]$ such that
\begin{equation}\label{3.1}
\langle w',Y(u,z_1)Y(v,z_2)w\rangle=\iota_{12}f(z_1,z_2).
\end{equation}
\noindent
(ii) {\bf Commutativity}: If $u,v,w,w',f$ are as in (i) then we also 
have
\begin{equation}\label{3.2}
\langle w',Y(v,z_2)Y(u,z_1)w\rangle =\iota_{21}f(z_1,z_2).
\end{equation}
(iii) {\bf Associativity}: In the same notation,
\begin{equation}\label{3.3}
\langle w',Y(Y(u,z_0)v,z_2)w\rangle=\iota_{20}f(z_0+z_2,z_2).
\end{equation}

\begin{lem}\label{l3.4} Let $V$ be a simple VOA  and 
$M$ a simple $V$-module. If 
$u_1,...,u_n$ are non-zero vectors in $V$ and 
$w_1,...,w_n$ are linearly independent vectors in $M,$ then
$$\sum_{i=1}^nY(u_i,z)w_i\neq 0.$$
\end{lem}

\noindent{\bf Proof:} Assume false. Fix $w'\in M'$ and
$v\in V.$ According to rationality (\ref{3.1}) there is 
a suitable $f\in \C(z_0,z_1)$ such that
$$\langle w',Y(v,z_1)\sum_{i=1}^nY(u_i,z_0)w_i\rangle=\iota_{10}f(z_1,z_0).$$
Because $\iota_{10}$ is injective, and since we are assuming that
$\sum_{i=1}^{n}Y(u_i,z)w_i=0,$ it follows that
$f(z_1,z_0)=0.$ Now use commutativity (\ref{3.3}) to see that
also
$$\langle w',\sum_{i=1}^nY(u_i,z_0)Y(v,z_1)w_i\rangle=0.$$
Since $w'\in M'$ is arbitrary we conclude that 
$$\sum_{i=1}^nY(u_i,z_0)Y(v,z_1)w_i=0.$$
Continuing in this way, we can obtain
\begin{equation}\label{3.4'}
\sum_{i=1}^nY(u_i,z_0)Y(v_1,z_1)\cdots Y(v_k,z_k)w_i=0
\end{equation}
for $k\geq 0$ and all $v_1,...,v_k\in V.$

If we let  $A$ be the associative algebra generated by
all component operators $u_m$ on $M$ for $u\in V$ and $m\in\Z,$ what 
(\ref{3.4'}) says is that
\begin{equation}\label{3.4}
\sum_{i=1}^nY(u_i,z)aw_i=0,\ \ \ {\rm for \ all}\ \ a\in A.
\end{equation}

Now there is no loss in assuming that 
the elements  $w_i$ are homogeneous, that is  $w_i\in M_{n_i}$ for some
$n_i.$  
If $A_0$ is the subalgebra of $A$ leaving each $M_m$ invariant it is
readily seen (eg. [IY], Lemma 2.1) that since 
$M$ is a simple $V$-module then $A_0$ induces the 
algebra $\sum_{i=1}^n\End\,M_{n_i}\subset \End M$ in its action on 
$\sum_{i=1}^nM_{n_i}.$
Then  we can choose $a\in A_0$ such that
$aw_1\ne 0,$ $aw_i=0$ for $i\geq 2.$  

The effect of this, together with
(\ref{3.4}) is, to allow us to assume that $n=1.$ In this case the Lemma is proved
in  [DL], so we are done.\ \ $\Box$
\bigskip

Now we can prove Theorem \ref{t2}. We are going to apply Lemma 
\ref{l3.4} with $M=V$ which is justified by our assumption that 
$V$ is a simple VOA. Set 
$$\Lambda=\{\chi\in\Irr(G)|V^{\chi}\ne 0\}.$$
We must establish that $\L=\Irr(G).$
Choose $\alpha,\beta\in\Lambda$ and let $\g\in\Irr(G)$ be a constituent of 
$\alpha\otimes\b.$ We are going to show that $\g\in\L.$

By assumption there are simple $G$-modules 
$W_{\alpha},$ $W_{\beta},$ $W_{\g}$ affording $\a,\b,\g$  respectively
such that $W_{\a}\subset V^{\alpha}, W_{\b}\subset V^{\beta},$
and $W_{\g}$ is (isomorphic to) a $G$-submodule of 
$W_{\alpha}\otimes W_{\beta}.$ So we can choose
non-zero vectors $u_1,...,u_n\in W_{\alpha}$ and
linearly independent vectors $w_1,...,w_n\in W_{\beta}$ with the 
property that
$\sum_{i}u_i\otimes w_i\in W_{\gamma}$ is non-zero.

By Lemma \ref{l3.4} we have $\sum_{i=1}^nY(u_i,z)w_i\ne 0,$ so
that there is $k\in\Z$ such that
$\sum_{i=1}^n(u_i)_kw_i\ne 0.$ Now by the previous paragraph we
have $\sum_{i=1}^n(u_i)_kw_i\in V^{\gamma},$ so $V^{\gamma}\ne 0$
and $\g\in\L,$ as desired.
 
We have shown that 
$\L$ contains all simple $G$-characters which 
are constituents of $\alpha\otimes\beta$ whenever  $\a,\b\in\L.$
On the other hand since $V$ is a faithful $G$-module there are
characters in $\L$ whose sum is a faithful character $\phi$ of $G.$

A famous theorem
of Burnside ([CR], Theorem 9.34) tells us that {\em every} 
simple character $\chi$ of $G$ lies in $\phi^{\otimes n}$ for 
some $n,$ so $\chi\in\L$  by the previous paragraph. This completes 
the proof of Theorem \ref{t2}. \ \ $\Box$ 

\begin{lem}\label{l3.4'} Let $G$ be a finite group and let $H,K$ 
be distinct subgroups of $G.$ There there is a simple $\C G$-module
$W$ such that the $H$-invariants $W^H$ and $K$-invariants $W^K$ are
distinct.
\end{lem}

\noindent{\bf Proof: } Without loss we may take $H\lneq K\leq G.$ 
Note that in the permutation module $\Ind_H^K(1)$ the $K$-invariants 
are {\em strictly} contained in the $H$-invariants, 
so the same must be true of some simple $G$-module contained in 
$\Ind_H^G(1)=\Ind_K^G\Ind_H^K(1).$\ \ $\Box$

Combining this and Theorem \ref{t2} yields
\begin{prop}\label{p3.3} Let $V$ be a simple VOA and $G$ a finite and faithful group of automorphisms of $V.$ Then 
the map $H\mapsto V^H$ is an injection from the set of subgroups of $G$
to the set of sub VOAs of $V$ which contains $V^G.$\ \ $\Box$
\end{prop}
\newpage

\section{A generational Result} 
\setcounter{equation}{0}
 
Let $V$ be a VOA and $M$ a $V$-module. For a subset $S\subset M,$
let $\< S\>$ denote 
the $V$-submodule of $M$ {\em generated} by $S,$ i.e., the 
smallest $V$-submodule of $M$ which contains $S.$ Obviously
$\< S\>$ is spanned by all of elements of the form 
$$v^1_{i_1}\cdots v_{i_k}^ks,\ \ v^i\in V, i_j\in \Z, s\in S,k\geq 0.$$

\begin{prop}\label{p4.1} $\langle S\rangle$ is spanned by all
elements of the form 
$$v_{n}s,\ \ \ v\in V, n\in \Z, s\in S.$$
\end{prop}

\noindent{\bf Proof:} We use another duality argument. If $U$ is  
the span of the elements $v_ns$ then we may decompose 
$\langle S\rangle=U\oplus X$ as a linear space where 
$X$ is a graded subspace of $\<S\>.$ 
Identify the restricted dual $X'$ of $X$  
with the annihilator of $U$ in $\< S\>'.$ 

Fix $w'\in X'$ and $w\in S.$ Then for $u,v\in V$ we 
obviously have
$$\langle w',Y(Y(u,z_0)v,z_2)w\rangle=0.$$
Associativity (\ref{3.3}) forces $f(z_0+z_2,z_2)=0,$ that
is $f(z_1,z_2)=0,$ whence rationality then implies that
$$\langle w',Y(u,z_1)Y(v,z_2)w\rangle =0.$$
Continuing in this way we find that
$$\langle w',v^1_{i_1}\cdots v^k_{i_k}w\rangle =0$$
for all $v^i\in V,$ $w\in S$ and $k\geq 2.$ This says precisely
that $\langle w',\< S\> \rangle=0,$ whence $w'=0.$ So
$X=0$ and the proposition is proved.\ \ $\Box$

\begin{coro}\label{c4.2} Let $V$ be a VOA 
and $M$ a simple $V$-module. Then for 
each $0\ne w\in M,$ $M$ is spanned as $V$-module by 
$v_nw$ for $v\in V,n\in\Z.$\ \ $\Box$
\end{coro}

\begin{rem} The result in Proposition \ref{p4.1} has also been obtained
independently by Li by a slightly different argument [L].
\end{rem}

These results will prove valuable in the rest of the paper.

Let $V$ be a simple VOA and $G$  a finite and faithful group 
of automorphisms of $V.$
\begin{th}\label{t4.3} 
For each linear (degree 1) character $\lambda$ of $G,$ 
$V^{\lambda}$ is a (non-zero) simple $V^G$-module. In particular,
$V^G$ is a simple VOA.
\end{th}

\noindent{\bf Proof:}  Observe that $V^{\l}$ is non-zero by Theorem \ref{t2}. 
Recall next that if $\lambda$ is a linear character
of $G$ and if $\chi\in\Irr(G),$ then $\lambda\chi\in\Irr(G)$ and
moreover $\lambda\chi=\lambda$ if, and only if, $\chi=1.$

Now pick $0\ne v\in V^{\l}.$ For any $u\in V^{\chi}$ we get 
$u_nv\in V^{\lambda\chi}$ for all $n\in\Z.$ But as $V$ is simple then 
 by Corollary \ref{c4.2} $V$ is spanned by all $u_nv$ for
$u\in V^{\chi},$ $\chi\in\Irr(G)$ and $n\in\Z.$ It follows that
$V^{\lambda}$ is necessarily 
spanned by all $u_nv$ for $u\in V^G,$ $n\in\Z,$ so that $V^{\lambda}$ is
a simple $V^G$-module. \ \ $\Box$
\newpage

\section{The case $G$ abelian}
\setcounter{equation}{0}

We complete the proof of Theorems \ref{tt1} and \ref{t3} in case $G$ is 
abelian in this 
section. So let $V$ be a simple VOA with a finite and faithful 
abelian group automorphisms
$G.$ According to Theorem \ref{t2} and Theorem \ref{t4.3} we have the 
decomposition
\begin{equation}\label{5.1}
V=\oplus_{\lambda\in\Irr(G)}V^{\lambda}
\end{equation}
and each $V^{\lambda}$ is non-zero simple $V^G$-module.

\begin{th}\label{t5.1} If $\lambda$ and $\mu$ are distinct
elements in $\Irr(G)$ then $V^{\lambda}$ and $V^{\mu}$ 
are inequivalent
$V^G$-modules.
\end{th}

\noindent{\bf Proof:} Otherwise there is an 
isomorphism of $V^G$-modules $\phi: V^{\lambda}\to V^{\mu}.$ Choose
$0\ne w\in V^{\lambda}$ and let $W$ be the $V$-submodule of
$V\oplus V$  generated by $(w,\phi(w)).$ By
Proposition  \ref{p4.1} $W$ is spanned by elements of the form
$(u_nw,u_n\phi(w))$ for $u\in V^{\nu}, \nu\in\Irr(G)$ and $n\in\Z.$

Now if  $u_nw\in V^{\lambda}$ and $u_nw\ne 0$ then necessarily
$u\in V^G,$ so that $(u_nw,u_n\phi(w))=(u_nw,\phi(u_nw)).$ This
makes it clear that $V^{\lambda}\oplus V^{\lambda}$ is not contained in
$W,$ so that $W\subsetneq V\oplus V.$ Hence $W\simeq V$ as $V$-modules
since $V$ is simple.

So projection of $W$ onto either factor in $V\oplus V$ is a $V$-module
isomorphism, hence so too is the map $u_nw\mapsto u_n\phi(w)$
of $V$ to itself. This latter is necessarily multiplication by a scalar
(Schur's lemma) since again $V$ is simple, and this
yields the contradiction that $\lambda=\mu.$ The theorem
is proved. \ \ $\Box$

Now $V^{\lambda}$ is both a $\C G$-module and a (simple) $V^G$-module. 
Furthermore if $v\in V^G$ then $g$ commutes with all component 
operators $v_n,$ $n\in\Z.$ Because of this it is straightforward to 
see that $V^{\lambda}$ is a simple module for the $G$-graded 
VOA $\C G\otimes V^G$ (see Remark \ref{r2.1}). So there is an 
identification $V^{\lambda}=M_{\lambda}\otimes V_{\lambda}$ for
a simple $V^G$-module $V_{\lambda}$ and simple $G$-module $M_{\lambda}$
affording the character $\lambda.$ 

This completes all parts of the proof of Theorem \ref{t3} for $G$
abelian. We turn to the proof of Theorem \ref{tt1}, which is now
easy.

If $W$ is a sub VOA of $V$ satisfying $V^G\subset W$ then $W$ is, 
in particular, a $V^G$-module. By Theorem \ref{t3} (for $G$ abelian)
we have $W=\sum_{\lambda\in\Lambda}V^{\lambda}$ for
some subset $\Lambda\subset \Irr(G).$  As $v_nw\in W\cap V^{\lambda\mu}$
whenever $v\in V^{\l},w\in v^{\mu}$ and $\l,\mu\in\L,$ it follows using
Lemma \ref{l3.4} that $\L$ 
is a subgroup of $\Irr(G).$ So if $H\leq G$ is the subgroup dual to $\L,$
i.e., $H=\{h\in G|\l(h)=1,\l\in\L\},$ then $W=V^H.$ Theorem \ref{tt1} 
now follows by duality for finite abelian group.
\newpage

\section{ Proof of Theorems 3 and 4}
\setcounter{equation}{0}
The first result will be a key point in our proofs of Theorems 3 and 4. We
assume that the following situation holds: 

$V$ is a simple VOA and $g$ an automorphism of $V$ of prime order $p;$
$M$ is a simple $V$-module such that $g\circ M$ is {\em not} isomorphic
to $M$ as $V$-modules.

\begin{th}\label{t6.1} With the above assumptions, $M$ is a simple
$V^{\<g\>}$-module.
\end{th}

\noindent{\bf Proof: } Since $g$ has prime order $p$ then the $V$-modules
$M_i=g^i\circ M, 0\leq i\leq p-1,$ (so that $M=M_0$) are pairwise inequivalent.
If $M$ has $L(0)$-grading $\coprod_{n\geq 0}M_{n+n_0}$ then
each $M_i$ has  a similar grading and we let $W_i=(M_i)_{n_0}$ be the
``top level'' of $M_i,$ i.e., the non-zero homogeneous subspace
of $M_i$ of least weight. The theory of the Zhu algebra $A=A(V)$
tells us that if $K_i$ is the kernel  of the action of $A$ on $W_i$
then $A/K_i\simeq \E(W_i).$ Furthermore if $K=\cap_{i=0}^{p-1}K_i$ then
$A/K\simeq \oplus_{i=0}^{p-1}\E(W_i).$

Next, from the definitions (\ref{2.12}) and (\ref{a2}) we see that
$g$ induces an algebra automorphism of $A.$  In fact we see that
$g$ maps $K_i$ to $K_{i+1}$ and so induces an automorphism of $A/K$ 
which permutes the summands $A/K_i.$  For $j=0,...,p-1$ let $A^j$ be the 
eigenspace of $g$ in $A$ with eigenvalue $\l^j$ ($\l=e^{2\pi i/p}$).
Then $A^j$ is the image of $V^j=\{v\in V|gv=\l^jv\}$ in $A$ and
$A=\oplus_{j=0}^{p-1}A^j.$ As a consequence we see that
$A/K=\oplus_{j=0}^{p-1}(A^j+K)/K$ is the decomposition of 
$A/K$ into eigenspaces of $g$ and 
$$A^j+K/K=\{(a,g\l^{-j}a,...,g^{p-1}\l^{-j(p-1)}a)|a\in A/K_1\}\subset \oplus_{i=1}^p\E(W_i).$$ 
In particular, $A^j+K/K$ acts irreducibly on each $W_i$ (this implies that
$(A^j+K/K)\cdot w=W_i$ for any non-zero $w\in W_i$).  

If $A_0=A(V^{\<g\>})=V^{\<g\>}/O(V^{\<g\>})$ then there is an algebra morphism
$A_0\to A$ defined in the obvious way. From the foregoing we conclude
that $A^0+K/K$ is the image of $A_0$ in $A/K.$ In particular, $A_0$ acts 
irreducibly on each $W_i.$ 

For convenience, set $M=M_1$ and $W=W_1.$  
Fix $0\ne w\in W,$ and let $M^{j}$ be the 
subspace of $M$ spanned by $u_nw$ for $u\in V^j$ and $n\in\Z.$
By Corollary \ref{c4.2} we get $M=\sum_{j=0}^{p-1}M^j.$

By the argument used to establish Proposition \ref{p4.1} we find that
$u_n M^j\subset M^{i+j}$ for $u\in M^i$ and $n\in \Z.$ In particular
each $M^j$ is a $V^{\<g\>}$-module. Since $A^j+K/K$ acts irreducibly on
$W,$ or equivalently, the operators $\{o(u)|u\in V^j\}$ (see 
Theorem \ref{t2.1} for the definition of $o(u)$) act
irreducibly on $W$ we see that $\{o(u)w |u\in V^j\}=W$ 
and $W\subset M^j.$ So $W\subset \cap M^j.$ But this latter intersection
is a $V$-module, so we find that $M=\cap_{j=0}^{p-1}M^j.$ 
This implies that $M=M^0$ is spanned by $w$ as a $V^{\<g\>}$-module for
any $0\ne w\in W.$

Finally, let $m\ne 0$ be a homogeneous element of $M.$ Let $M^j$ be as 
before, but now with $w$ replaced by $m.$ Then again each $M^j$ is a 
$V^{\<g\>}$-module and $u_n M^j\subset M^{i+j}$ for 
$u\in V^i$ and $n\in\Z.$ 
By irreducibility of $M$ there
exists $i$ such that $W\cap M^i\ne 0.$ 
Since $W$ is a simple $A_0$-module we see that $W\subset M^i.$ 
As $A^j+K/K$ acts irreducibly on $W$ we again get 
$W\subset M^{i+j}$ for all $j.$ In
particular, $W\subset M^0.$ Since $M$ is spanned by $0\ne w\in W$ as 
$V^{\<g\>}$-module, this forces $M^0=M.$

Thus for each $0\ne m\in M,$ $M$ is spanned by the elements 
$u_nm$ for $u\in V^{\<g\>}$ and $n\in\Z.$ So $M$ is a simple
$V^{\<g\>}$-module. \ \ $\Box$
\bigskip

We turn our attention to the proof of Theorem \ref{t3}. We proceed by 
induction on the order of $G.$ As $G$ is solvable there is a 
normal subgroup $H\vartriangleleft G$ of prime index $p$ in $G.$ Inductively,
we may assume that there is a decomposition of $V$ into $V^H$-modules of the
form 
\begin{equation}\label{6.1}
V=\oplus_{\chi\in\Irr(H)}M_{\chi}\otimes V_{\chi}
\end{equation}
where $M_{\chi}$ is a simple $\C H$-module affording the character $\chi$ 
of $H,$ where $V_{\chi}$ is a simple $V^H$-module,  where 
$V_{\chi}\not\simeq V_{\mu}$ for distinct $\chi,\mu$ and
$M_{\chi}\otimes V_{\chi}$ is a simple module for the $H$-graded 
VOA $\C H\otimes V^H.$

Before proceeding we recall the following well-known result
(cf. [CR], (11.1) and (11.47)): 
\begin{lem}\label{l6.2} Let $\chi\in \Irr(H)$ and set $\psi=\Ind_H^G(\chi).$
One of the following holds:

(i) $\psi\in\Irr(G).$ In this case $\Res_H^G(\psi)=\chi_1+\cdots+ \c_p$
where $\c=\c_1,...,\c_p$ are the distinct $G$-conjugates of $\c$ in $\Irr(H).$

(ii) $\psi\not\in \Irr(G).$ In this case $\psi=\alpha_1+\cdots+\a_p$ where 
$\a_1,...,\a_p$ are the distinct characters in $\Irr(G)$ whose
restriction to $H$ is $\c.$ We have $\a_i=\a_1\l_i,$ $1\leq i\leq p,$ where
$\l_1,...,\l_p$ are the characters of $G/H$ lifted to $G.$
\end{lem}

Now since $H\vartriangleleft G$ then $G$ acts  on $\Irr(H)$ and hence if 
$g\in G$  we get $g(V^{\c})=V^{g\c}$ where $V^{\chi}=M_{\c}\otimes V_{\c}.$

Fix $\c\in\Irr(H)$ and suppose to begin with that case (i) of Lemma
\ref{l6.2} holds. With the notation of the lemma
we get $g\cdot \c\ne \c$ for $g\in G\backslash H,$ so $V_{\c}$ and
$g(V_{\c})$ are inequivalent simple $V^H$-modules. As $V^H$ is a simple 
VOA (Theorem \ref{t4.3}) we may apply Theorem \ref{6.1} to conclude that
$V_{\c}$ is a simple $V^G$-module. Indeed $V_{\c_1},...,V_{\c_p}$ are
equivalent, simple $V^G$-modules.

From (\ref{6.1}) we now find that there is an isomorphism of $V^G$-modules
\begin{equation}\label{6.2}
\oplus_{i=1}^pM_{\chi_i}\otimes V_{\chi_i}\simeq M_{\psi}\otimes V_{\psi}.
\end{equation}
Indeed this follows since $V_{\psi}=\QRes_H^G(V_{\c})$ and 
$M_{\psi}=\Ind_H^G(M_{\c}).$ As each $M_{\c_i}\otimes V_{\c_i}$ is a 
simple $\C H\otimes V^H$-module we easily find that $M_{\psi}\otimes V_{\psi}$
is a simple $\C G\otimes V^G$-module. 

Now assume that case (ii) of Lemma \ref{l6.2} holds. This is the case in which $\c\in\Irr(H)$ is $G$-invariant, that is $g\cdot\c=\c$ for all $g\in G.$ Thus
$G$ leaves $V^{\chi}$-invariant. Thus if we fix 
$g\in G\backslash H$ then $V_{\c}\simeq g(V_{\c})$ as $V^H$-modules.
Fix such an isomorphism $\a: g(V_{\c})\to V_{\c}.$ Thus 
we have
$$\a Y(v,z)\a^{-1}w=Y(v,z)w,\ \ v\in V^H, w\in V_{\c}.$$
Setting $\b=\a g,$ this yields $\b: V_{\c}\to V_{\c}$ and
\begin{equation}\label{6.3}
\b Y(v,z)\b^{-1}w=Y(gv,z)w,\ \ v\in V^H, w\in V_{\c}.
\end{equation}

As $g^p\in H$ we see from (\ref{6.3}) that $\b^p$ induces a $V^H$-isomorphism
of $V_{\c}$ onto itself. By Schur's lemma $\beta^p$ is a scalar in its action
 on $V_{\c},$ and we then see that the eigenvalues of $\beta$ on $V_{\c}$
are precisely the form $\rho^ic$ for some fixed nonzero scalar $c$ and
$0\leq i\leq p-1,$ where $\rho$ is a primitive pth root of unity.

Again from (\ref{6.3}), $\b$ commutes with action of the vertex operator
$Y(v,z)$ for $v\in V^G,$ so each eigenspace of $\b$ on $V_{\c}$ is a 
$V^G$-module.

We contend that these $p$ $V^G$-submodules are in fact simple.
Indeed the assertion follows as in the proof of Theorem \ref{t4.3}.

Now after Lemma \ref{l6.2} (ii) we must have $V^{\c}=\oplus_{i=1}^pV^{\a_i}.$
As $M_{\c}=\Res_H^G(M_{\a_i})$ for each $i,$ we have 
$V^{\chi}=\oplus_{i=1}^pM_{\a_i}\otimes V_{\a_i}$ where by definition 
$V_{\a_i}$ is the $\rho^ic$-eigenspace of $\b$ on $V_{\c}.$

Finally, it is readily seen that since $\C H\otimes V^H$ is irreducible on 
$V^{\c}=M_{\c}\otimes V_{\c}$ and since $V^G$ is irreducible on $V_{\a_i},$
then also $\C G\otimes V^G$ is irreducible on 
$V^{\a_i}=M_{\a_i}\otimes V_{\a_i}.$

Now part (i) of Theorem \ref{t3} follows for $G,$ and hence is proved.
We turn our attention to part (ii). 

We must prove that if $\a,\b\in\Irr(G)$ are such that 
$V_{\a}\simeq V_{\b}$ as $V^G$-modules then $\a=\b.$ 
Let $\phi: V_{\a}\to V_{\b}$ be a $V^G$-isomorphism and suppose first that
both $\Res_H^G(\a)=\chi$ and $\Res_H^G(\b)=\psi$ lie in $\Irr(H).$ Thus we are in case (ii) of
Lemma \ref{l6.2}, where we have already established that 
$V_{\chi}=\oplus_{i=1}^{p}V_{\a_i}$
and $V_{\psi}=\oplus_{i=1}^{p}V_{\b_i}$
 where $\a=\a_1,...,\a_p,$ $\b=\b_1,...,\b_p,$
$\a_i=\l^i\a_1$ and $\b_i=\l^i\b_1$ as in Lemma \ref{l6.2} (ii). 

Now we have $V^H=\oplus_{i=1}^{p}V^{\l_i}$ and 
$Y(v,z)u\in V_{\a_i}[[z,z^{-1}]],$ 
$Y(v,z)w\in V_{\b_i}[[z,z^{-1}]]$ 
for $v\in\l_i, u\in V_{\a},w\in V_{\b}.$ 
So if we choose $0\ne w\in V_{\a}$ and consider the $V^H$-submodule 
generated by 
$(w,\phi(w))$ we get $\a=\b$ via the argument used to prove Theorem 
\ref{t5.1}.

Next assume that $\Res_H^G(\a)\not\in\Irr(H),$ and similarly
$\Res_H^G(\b)\not\in \Irr(H).$ So there are $\chi,\psi\in \Irr(H)$
such that $\a=\Ind_H^G(\chi), \b=\Ind_H^G(\psi).$
If $\chi$ and $\psi$ are $G$-conjugates then certainly $\a=\b,$
so we may assume that this is not the case. Thus the distinct $G$-conjugates of
$\chi$ and $\psi$ provide us with $2p$ distinct elements of $\Irr(H),$ call
them $\ch=\ch_1,...,\ch_p,$ $\psi=\psi_1,...,\psi_p.$

Let $W_i,X_i$ $1\leq i\leq p$ be the ``top levels'' of the simple
$V^H$-modules $V_{\chi_i}, V_{\psi_i}$ respectively. By induction
they are inequivalent $V^H$-modules, so they afford inequivalent
simple $A(V^H)$-modules. Thus this latter algebra is represented
on $\oplus_{i=1}^pW_i\oplus\left(\oplus_{i=1}^pX_i\right)$ as
a quotient of $A(V^H),$ say $\bar A\simeq \oplus_{i=1}^p\End(W_i)\oplus
\left(\oplus_{i=1}^p\End(X_i)\right).$ If we now consider the action of 
$g\in G\backslash H$ on this space, we
find as in the proof of Theorem \ref{t6.1} (with $V$ in that theorem replaced
by our present $V^H$) that $g$ acts on
$\bar A$ in such a way that $\bar A^{\<g\>}$ is the image of 
$A(V^G)\to A(V^H)\to \bar A$ under the natural homomorphism, and
that $\bar A^{\<g\>}$ is isomorphic to $\End(W_1)\oplus\End(X_1)$
as algebras.

The point is that it follows from these considerations that
$A(V^G)$ is represented irreducibly on $W_1$ and $X_1,$ and that
these representations are inequivalent. Thus by Zhu's basic Theorem \ref{t2.1}
it follows that $V_{\ch_1}$ and $V_{\psi_1}$  $-$ alias 
$V_{\alpha}$ and $V_{\b}$ $-$ are inequivalent $V^G$-modules. 

Thus finally we may assume that $\Res_H^G(\a)=\chi\in \Irr(H)$
and $\Res_H^G(\b)\not\in \Irr(H).$  Then 
$V_{\chi}=\oplus_{i=1}^pV_{\a_i}$ where $\a=\a_1,\a_2, ...,\a_p$
and $\a_i=\l^i\a$ as in the first case. Let $\psi\in \Irr(H)$ such
that $\beta=\Ind_H^G(\psi).$ Then $g^iV_{\psi}\simeq V_{\psi}$ as
$V^G$-modules. 
Consider the
$V^H$-module $N=\sum_{i\geq 0}g^iV_{\psi}.$ Let  $m$ be the order of
$g$ on $N.$ Then $N=\sum_{i=0}^{m-1}N^{\mu_i}$ where
$\mu_0,...,\mu_{m-1}$ are distinct characters for $\<g\>$ on $N.$ 
Set $W=V_{\chi}\oplus N.$ 
Also consider the $V^H$-submodule $M$ of $W$ generated by 
$(w,\phi(w),g\phi(w),...,g^{m-1}\phi(w))$ where $0\ne w\in V_{\a}.$
Again by Proposition \ref{p4.1}, 
$M$ is spanned by 
\begin{equation}\label{addd}
(u_nw,u_n\phi(w),u_ng\phi(w),...,u_ng^{m-1}\phi(w))
\end{equation}
for $u\in V^H.$  Note that the element in (\ref{addd}) lies in
$V_{\alpha_i}\oplus N^{\l^i}$ for $u\in V^{\l^i}.$
Thus $V_{\alpha}\nsubseteq M$ and therefore
$M\cap V_{\chi}=0$ since $V_{\chi}$ is a simple $V^H$-module. 
By our inductive hypothesis, $V_{\chi}$ is inequivalent to any
simple submodule of $N.$ This implies that $M$ does not have
any composition factor isomorphic to $V_{\chi}.$
One the other hand the projection of $M$ to $V_{\chi}$ 
is a $V^H$-epimorphism.
This yields a contradiction and 
the proof of Theorem \ref{t3} (ii) is complete.
\smallskip

We will demonstrate Theorem \ref{t4} as a consequence of Theorem \ref{t3}
and Frobenius reciprocity. To begin with, the 
categories ${\cal V}(G)$ for $G$ solvable are
semi-simple by Theorem \ref{t3}, as of course is the 
category $G$-Mod. Moreover the 
induction functor $\Ind_H^G$ is additive, as is the functor $\QRes_H^G.$
In view of this it is sufficient to prove the following.
If $H\leq G,$ $\psi\in \Irr(H)$ and $\Ind_H^G(\psi)=\sum_{i}n_i\chi_i$
with $\c_i$ the distinct elements of $\Irr(G)$ and $n_i\geq 0;$ and
if $V_{\psi}, V_{\chi_i}$ are the corresponding simple $V^H$- respectively
$V^G$-modules as in Theorem \ref{t3}, then $\QRes_H^G(V_{\psi})=\oplus_{i}n_iV_{\c_i}.$

For each $i,$ the multiplicity of $\psi$ in $\Res_H^G(\c_i)$ is $n_i$ by
Frobenius reciprocity. Thus the $\C G\otimes V^{G}$-module
$V^{\c_i}=M_{\c_i}\otimes V_{\c_i}$ contains $n_i$ copies of the linear
space $M_{\psi}\otimes V_{\c_i}.$ From this it follows, with an obvious
notation, that the $V^H$-module $V^{\psi}=M_{\psi}\otimes V_{\psi}$ can
be written as $M_{\psi}\otimes(\oplus_{i}n_iV_{\c_i}).$ Our assertion 
follows.
\newpage

\section{Completion of proof of Theorem 1}
\setcounter{equation}{0}

After the results of Section 5, in proving Theorem \ref{tt1} we may take 
$G$ to be a dihedral group of order $2N,$ say. We again use induction on the order of $G,$ the result being already true for abelian groups.

Let $H\vartriangleleft G$ be a  cyclic subgroup of $G$ of order $N.$ So we 
have decomposition into $V^H$-modules as usual:
$$V=\oplus_{\l\in\Irr(H)}V^{\l}$$
with each $V^{\l}\ne 0.$

For simplicity we shall take $N$ odd in what follows. The case in 
which $N$ is even is dealt with in the same way with only superficial 
changes in the proof. With $N$ odd, $G$ has
just two linear characters $1$ and $\sigma,$ say, and $(N-1)/2$ simple
characters of degree 2, call them $\psi_1,...,\psi_{\frac{N-1}{2}}.$ We
may choose notation so that $\psi_i=\Ind_H^G(\l_i)
=\l_i+\bar\l_i,$ $1\leq i\leq (N-1)/2,$ where $\l_i\in \Irr(H)$ is the
character on which a fixed generator $h$ of $H$ takes value $\rho^i,$
for some fixed primitive $N$th root of unity $\rho.$ Thus in terms of 
$V^G$-modules we have a decomposition 
\begin{equation}\label{7.1}
V=V^G\oplus V^{\sigma}\oplus \left(\oplus_{i=1}^{(N-1)/2}M_{\psi_i}\otimes V_{\psi_i}\right)
\end{equation}
in earlier notation. Moreover $M_{\psi}\otimes V_{\psi_i}=V^{\psi_i}
=V^{\l_i}\oplus V^{\bar\l_i}.$

Now let $W$ be a VOA satisfying $V^G\subset W\subset V.$ We must prove
that $W=V^K$ for some subgroup $K\leq G.$ Suppose first that 
$V^{\sigma}\subset W.$ In this case $W$ contains $V^G\oplus V^{\sigma}=V^H$ 
and as $H$ is abelian we are done by the results of Section 5.

From now on, then, we assume that $V^{\sigma}\nsubseteq W.$ As $V^{\sigma}$ is a simple $V^G$-module (Theorem \ref{t5.1}) then $V^{\sigma}\cap W=0.$

Next we claim that $V^{\l}\cap W=0$ for all $\l\in\Irr(H).$ For if not,
using Lemma \ref{l3.4} (and the discussion that follows it) one finds by 
iterating suitable operator $u_n$ applied to $w$ for $u\in V^{\l}$ and 
$0\ne w\in V^{\l}\cap W,$ $n\in\Z,$ that one obtains an element
in $V^H\backslash V^G,$ which is impossible as $V^{\sigma}\cap W=0.$ 
So indeed $V^{\l}\cap W=0$ for all $\l.$

On the other hand if $V^{\psi_i}\cap W=0$ for all $i$ then $W=V^G$ and we
are done. So we may assume the existence of an index $i,$ $1\leq i\leq (N-1)/2,$ such that $V^{\psi_i}\cap W$ is a simple $V^G$-module (isomorphic to 
$V_{\psi_i}$).

Until further notice we assume that this index $i$ may be chosen that 
$\rho^i$ is a primitive $N$th root of unity. In this case it will make 
no difference if we take $i=1,$ so we assume this. As $V^{\psi_1}\cap W\simeq 
V_{\psi_1}$ and $W\cap V^{\l_1}=W\cap V^{\bar \l_1}=0$ it follows that
there is a linear isomorphism 
$$\phi: V^{\l_1}\to V^{\bar \l_1}$$
such that $V^{\psi_1}\cap W=\{(v,\phi(v))|v\in V^{\l_1}\}.$ Then 
in fact $\phi$ is an isomorphism of $V^G$-modules.

Now let $t$ be an involution in the dihedral group $G.$ Then $t$ inverts
$h,$ $tht=h^{-1},$ so that as an operator on $V,$ $t$ interchanges $V^{\l_1}$
and $V^{\bar \l_1}$ and is again an isomorphism of $V^G$-modules. Then 
$t\phi: V^{\l_1}\to V^{\l_1}$ is a $V^G$-automorphism of the simple
$V^G$-module $V^{\l_1},$ so that $t\phi$ operates on $V^{\l_1}$ as a 
scalar $c_1,$ say, by Schur's Lemma.

As $W$ is a VOA then $Y(v+\phi(v),z)(v+\phi(v))$ again lies in $W[[z,z^{-1}]]$
for $v\in V^{\l}$ and is non-zero for $v\ne 0$ by Lemma \ref{l3.4}. More 
precisely there is $n\in\Z$ such that 
$v_nv\ne 0$ if $v\ne 0.$ Note that $v_nv\in V^{\l_2}.$ As 
$Y(v,z)\phi(v)+Y(\phi(v),z)v$ lies in $V^G[[z,z^{-1}]]$ then
$0\ne v_nv+\phi(v)_n\phi(v)\in V^{\psi_2}\cap W.$

Now we can proceed as before: extend the range and the domain of $\phi$ by 
defining $\phi:$ $V^{\l_2}\to V^{\bar\l_2}$ in such a way that 
$V^{\psi_2}\cap W=\{v+\phi(v)|v\in V^{\l_2}\}.$ Again we find that 
$t\phi$ operators on $V^{\l_2}$ as a scalar $c_2.$ Continuing in this way 
we can construct isomorphisms of $V^G$-modules 
$\phi: V^{\l_i}\to V^{\bar \l_i},$
$1\leq i\leq (N-1)/2,$ such that $V^{\psi_i}\cap W=\{v+\phi(v)|v\in V^{\l_i}\}$
and $t\phi$ operates on $V^{\l_i}$ as a scalar $c_i.$ We may further
define $\phi: V^{\bar\l_i}\to V^{\l_i}$ so that
$\phi^2=id.$ Then also $t\phi$ operates on $V^{\bar\l_i}$ as multiplication by
$c_i^{-1}.$ 

Consider $Y(u+\phi(u),z)(v+\phi(v))$ for $u\in V^{\l_i}, v\in V^{\l_j}.$ Using
Lemma \ref{l3.4} as before, we find that we must  have, since
$u_nv\in V^{\l_{i}\l_j},$ $\phi(u)_n\phi(v)\in V^{\bar\l_i\bar\l_j},$ that
\begin{equation}\label{7.2}
\phi(u_nv)=\phi(u)_n\phi(v).
\end{equation}
Applying $t$ to both sides yields that for $i+j\leq (N-1)/2$ we have
$$c_{i+j}u_nv=t\phi(u_nv)=t(\phi(u)_n\phi(v))=(t\phi(u))_nt\phi(v)
=c_ic_ju_nv.$$
Choosing as we may $u_nv\ne 0,$ we conclude that $c_{i+j}=c_ic_j$
for $1\leq i,j,i+j\leq (N-1)/2.$ 

Hence 
\begin{equation}\label{7.3}
c_i=c_1^i,\ \ 1\leq i\leq (N-1)/2.
\end{equation}
If we take  $i=(N-1)/2,j=1$ in (\ref{7.2}) then we also get in the same way
that
\begin{equation}\label{7.4}
c_{\frac{N-1}{2}}^{-1}=c_{\frac{N-1}{2}}c_1.
\end{equation}
Combine (\ref{7.3}) and (\ref{7.4}) to see that 
\begin{equation}\label{7.5}
c_1^N=1.
\end{equation}
From (\ref{7.5}) we conclude that there is an integer $k$ such that $c_1=\rho^k.$ Moreover $t\phi$ operates on $V^{\l_i}$ as the scalar $\rho^{ki},$
and on $V^{\bar\l_i}$ as $\bar\rho^{ki}.$

By construction, a generator $h$ of $H$ acts on $V^{\l_i}$ as $\rho^i,$ so
$t\phi$  
acts on each $V^{\l_i}$ and each $V^{\bar\l_i}$ in the same way that $h^k$
does, and $\phi$ acts on these spaces as $th^k$ does.

However for $1\leq i\leq (N-1)/2$ we have 
$V^{\psi_i}\cap W=\{v+\phi(v)|v\in V^{\l_i}\}$ is the space of 
$\phi$-invariants  in $V^{\psi_i}.$ From this we conclude that the space 
of $th^k$-invariants in $\oplus_{i=1}^{\frac{N-1}{2}}V^{\psi_i}$ is
precisely the intersection of $W$ with this space.

Finally, since $th^k$ acts as $-1$ on $V^{\sigma},$ we can conclude from
(\ref{7.1}) that $W$ is precisely the space of $th^k$-invariants of $V.$ 
So the Theorem holds in this case.

Now we must deal with the possibility that for no index $i$ is it the case that
$V^{\psi_i}\cap W\ne 0$ and $\rho^i$ is a primitive $N$th root. Then 
using Lemma \ref{l3.4} we find that the set of indices $i,$
$1\leq i\leq (N-1)/2,$ for which $V^{\psi_i}\cap W\ne 0$ is such that
the corresponding set $\{\l_i,\bar\l_i\}$ together 1 is a subgroup
$\L$ of $\Irr(H).$ Let $L$ be the dual group i.e., the intersection 
(in $H$) of the kernels of $\l_i.$ Then 
$$V^L=V^{G}\oplus V^{\sigma}\oplus_{\l_i\in \L}V^{\l_i}\oplus V^{\bar\l_i}.$$

This reduces us to the last case since $G/L$ is a faithful abelian or dihedral
group of opertors on the simple VOA $V^L.$ So in this case $W$ is equal
to $V^K$ for some dihedral subgroup $K$ of $G$ containing $L$ with index 2.

This completes our discussion of Theorem \ref{tt1}.

\end{document}